\documentclass[12pt,preprint]{aastex}
\begin{document}
\title{Phenomenology of turbulent dynamo growth and saturation}

\author{Rodion Stepanov}
\affil{Institute of Continuous Media Mechanics, Korolyov 1, 614013
Perm, Russia} \email{rodion@icmm.ru}

\and

\author{Franck Plunian}
\affil{Laboratoire de G\'eophysique Interne et Tectonophysique,
Universit\'e Joseph Fourier, CNRS, Maison des G\'eosciences, B.P.
53, 38041 Grenoble Cedex 9, France} \email{Franck.Plunian@ujf-grenoble.fr}

\begin{abstract}
With a non local shell model of magnetohydrodynamic turbulence we investigate numerically the turbulent dynamo action for low and high magnetic Prandtl numbers ($Pm$). The results obtained in the kinematic regime and along the way to dynamo saturation are understood in terms of a phenomenological approach based on the local ($Pm\ll 1$)
or non local ($Pm\gg 1$) nature of the energy transfers. In both cases the magnetic energy grows at small scale and
saturates as an inverse `` cascade ''.

\end{abstract}


\section{Introduction}
Dynamo action is generally believed to be at the origin of the magnetic field in most astrophysical objects. The conducting fluid which produces the magnetic field, is generally strongly turbulent ($Re \gg 1$) with an extended inertial range.
In addition as it is electrically conducting, the fluid is characterized by the magnetic Prandtl number $Pm=\nu / \eta$, where $\nu$ and $\eta$ are the fluid viscosity and magnetic diffusivity. In a liquid metal as in planetary cores or stellar convective zones $Pm \ll 1$, while in the warm interstellar medium, coronal and cluster plasmas $Pm\gg 1$ \cite{Schekochihin02a}.
In direct numerical simulations both conditions,  $Re \gg 1$ and $Pm$ much different from unity,
 are still out of reach of the present day computers. An alternative is to use shell models.

Hydrodynamical shell models are a rough approximation of the Navier-Stokes equations formulated on a discrete set of real wave numbers $k_n=\lambda^n$, corresponding to a Fourier space divided into shells of logarithmic width $\lambda$. They involve (complex) scalar quantities reminiscent of the velocity Fourier components of an isotropic flow.
The model satisfies the conservation of both kinetic energy and kinetic helicity when the viscosity is set to zero. It leads to the resolution of a system of ordinary differential equations, requiring much less computing power than the direct numerical simulations of the original equations. It is then possible to run a shell model for realistic values of the viscosity such as $\nu=10^{-6} m^2\cdot s^{-1}$ (for a review on shell models, see Biferale (2003) and references therein).

The results obtained with shell models are in general agreement with the Kolmogorov phenomenology including intermittency \cite{Leveque97}. Firstly at any scale $k^{-1}$ lying in the inertial range, the flux rate of kinetic energy is equal to the injection rate of kinetic energy at the forcing scale. This writes $ku^3(k)\sim \varepsilon$, leading to $u(k)\sim \varepsilon^{1/3}k^{-1/3}$ and to the Kolmogorov spectrum $E(k)\sim \varepsilon^{2/3}k^{-5/3}$.
Secondly, the inertial range is found to extend to the viscous scale $k_{\nu}^{-1}$ which can be evaluated saying that at this scale the turn over time $k^{-1}u^{-1}$ compares with the viscous time $\nu^{-1}k^{-2}$, leading to $k_{\nu}\sim \varepsilon^{1/4}\nu^{-3/4}$.
Thirdly at low viscosity some deviation of the Kolmogorov power scaling is found  due to intermittency. The power scaling of the statistical moments are given by $<|u(k)|>^p\sim k^{-\zeta_p}$
with scaling exponents $\zeta_p$ deviating from Kolmogrov's mean field theory $\zeta_p=p/3$. The value of these anomalous scaling exponents given by the shell models compare well with those found in experiments \cite{Leveque97}.
Therefore shell models appear to be a useful tool to study fully developed turbulence at high Reynolds numbers.
They are also used in their MHD version to tackle astrophysical issues \cite{Frick06,Buchlin07,Galtier07}.

In a previous paper \cite{Plunian07}, denoted after by PS07, we introduced a non local shell model which is not only in agreement with the previous features of turbulence but in addition gives an appropriate slope of the infrared spectrum in freely decaying turbulence (which is not the case of the local shell models described in Biferale 2003).
The magnetohydrodynamic (MHD) version of this model, also introduced in PS07, permits to study turbulent dynamo action for arbitrary low or high values of $Pm$. In PS07 we calculated the energy transfer functions of the MHD system in a (statistically stationary) saturated state. We found that for $Pm \le 1$ the energy transfers are mainly local, eventually strengthening our previous results obtained with a local shell model of MHD turbulence \cite{Stepanov06}. For $Pm \gg 1$ the dominant transfers are also mainly local except the ones from the flow scales lying in the inertial range to the magnetic scales smaller than the viscous scale. In that case the use of a non local model is definitely necessary.

Here our goal is to describe the transient properties of the turbulent dynamo, namely the kinematic regime during which the magnetic energy grows exponentially,
and the route to saturation, starting when the kinematic regime stops and ending when a (statistically stationary) saturated state is reached.
For that
it is not sufficient to make time averages as we did in PS07 when studying the saturated dynamo states. Indeed the transient properties we are interested in are, by definition, not statistically stationary.
Then instead of time averages we have made ensemble averages over $M=12\cdot10^3$ independent realisations. The initial conditions have the same energy but random phases. In addition the forcing has also a random phase in time (therefore different from one realisation to the other). Then the results between two realisations are different. Every $t_0=10^{-2}$ unit of time we freeze the calculation. For each shell we average the magnetic and kinetic energy over the $M$ realisations without modifying the phases of the magnetic, kinetic and forcing quantities in order to keep the randomness between every realisations. All the calculation were done for $N=60$ shells, corresponding to more than 12 decades. Before presenting the results (sections \ref{lowpm} and \ref{highpm} for low and high $Pm$) we briefly describe in the next section the shell model (previously introduced in PS07).

\section{The non local shell model}
\label{model}
The model is defined by the following set of equations
\begin{eqnarray}
\dot{U}_n &=& i k_n \left[Q_n(U,U,a)-Q_n(B,B,a)\right]
- \nu k_n^2 U_n + F_n, \label{eq_u} \\
\dot{B}_n &=& i k_n \left[Q_n(U,B,b)-Q_n(B,U,b)\right]
- \eta k_n^2 B_n, \label{eq_b}
\end{eqnarray}
where
\begin{eqnarray}
 Q_n(X,Y,c)= \sum^{N}_{m=1}T_m [ c_m^1 X_{n+m}^*Y_{n+m+1}
 +  c_m^2 X_{n-m}^* Y_{n+1}
+ c_m^3 X_{n-m-1} Y_{n-1} ] \label{Qn}
\end{eqnarray}
represents the non linear transfer rates and
$F_n$ the turbulence forcing applied at shell $n=0$. As explained in PS07 an optimum shell spacing is the golden number $\lambda = (1 + \sqrt{5}) / 2$.

For $N=1$ in (\ref{Qn}), we recognize the local Sabra model \cite{Lvov98}. The additional non-local interactions for $N \ge 2$ correspond to all other possible triad interactions except the ones involving two identical scales.
Expressions for the kinetic energy and helicity $E_U$ and $H_U$, magnetic energy and helicity $E_B$ and $H_B$, and cross helicity $H_C$ are given by
\begin{eqnarray}
E_U = \sum_{n}E_U(n), \; E_U(n)=\frac{1}{2} |U_n|^2, &&
    H_U = \sum_{n}H_U(n), \; H_U(n)=\frac{1}{2} (-1)^n k_n |U_n|^2 \label{kinetic}\\
E_B = \sum_{n}E_B(n), \; E_B(n)=\frac{1}{2} |B_n|^2,&&
    H_B = \sum_{n}H_B(n), \; H_B(n)=\frac{1}{2} (-1)^n k_n^{-1} |B_n|^2 \label{magnetic}\\&&
    H_C = \sum_{n}H_C(n), \; H_C(n)=\frac{1}{2}  (U_n B_n^* + B_n U_n^*).\label{cross helicity}
\end{eqnarray}

In the inviscid and non-resistive limit ($\nu=\eta=0$), the total energy
$E=E_U+E_B$, magnetic helicity and cross helicity must be conserved ($\dot{E}=\dot{H}_B=\dot{H}_C=0$).
This implies the following expression for the coefficients $a_m^i$ and $b_m^i$:
\begin{eqnarray}
    a_m^1 = k_{m} + k_{m+1} \quad \quad&
    a_m^2 = \frac{-k_{m+1} - (-1)^m}{k_{m}} \quad \quad&
    a_m^3 =   \frac{k_{m} - (-1)^m}{k_{m+1}}\nonumber\\
    b_m^1 = (-1)^{m+1}&b_m^2 = 1&b_m^3 =  -1.
    \label{coefficients}
\end{eqnarray}
In the case of pure hydrodynamic turbulence (without magnetic field), the coefficients $a_m^i$
derived from the kinetic energy and helicity conservations ($\dot{E_U}=\dot{H}_U=0$), would lead to the same expression as (\ref{coefficients}).
The coefficients $T_m$ are free parameters depending on $m$ only, that we
choose of the form $T_m= k_{m-1}^{\alpha}/\lambda(\lambda +1)$.
Further characteristics of the model and results can be found in PS07.
In particular the role of the non locality parameter $\alpha$ has been investigated. It is indeed the only free parameter left in the model that can not be theoretically constrained.

In PS07 we found that taking $\alpha= -5/2$ permits to describe accurately the infrared hydrodynamic spectrum in freely decaying turbulence.
Keeping this value of $\alpha$ in the MHD system for $Pm\le 1$, we found that the dominant energy transfers are mainly local.

On the other hand, for $Pm \gg 1$, in order to describe properly the non local energy transfer from the flow scales lying in the inertial range to the magnetic scales smaller than the viscous scale, the results obtained in PS07 suggest to take a parameter $\alpha$ corresponding to stronger non local transfers than those obtained for $\alpha= -5/2$. A phenomenological justification of the choice $\alpha=-1$ will be given below. It can be also understood as a way to mimic in our (isotropic) model the strongly anisotropic small scale magnetic turbulence. It is worth mentioning here that taking $\alpha=-1$ does not affect too much the energy transfers involving the magnetic scales lying in the inertial range, which remain mainly local.

\section{Low $Pm$ dynamo action}
\label{lowpm}
\subsection{Kinematic regime}
\begin{figure}
\begin{tabular}{@{\hspace{4cm}}c@{\hspace{2.5cm}}c@{\hspace{-6.5cm}}c@{}}
                                  &                & \includegraphics[width=1\textwidth]{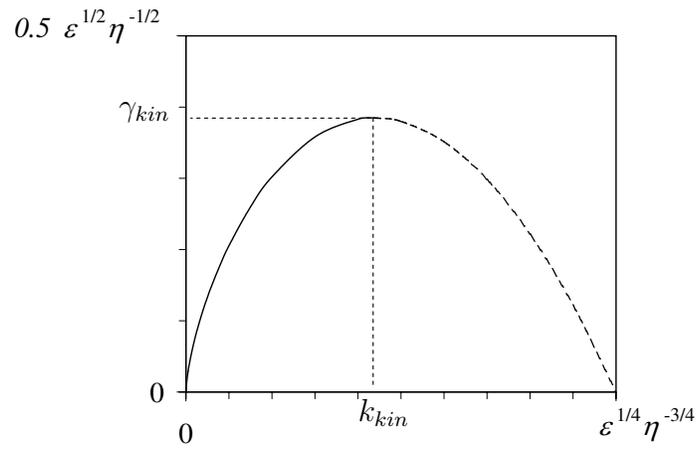} \\*[-6.5cm]
    \raisebox{6cm}{$\gamma_{kin}$}&                &                                                \\*[-2.5cm]
                                  &    $k_{kin}$  &                           \\*[0.5cm]
  \end{tabular}
\caption{Growthrate spectrum $\gamma(k)$ produced by $u(k)$ lying in the inertial range. }
\label{gamma}
\end{figure}

At low $Pm$, as shown in PS07, the energy transfers responsible for the generation of the magnetic from the kinetic energy are mainly local and the flow scales producing the largest magnetic growthrate lie in the inertial range. A phenomenological way to understand it is to write the induction equation at wave number $k$, in the form
\begin{equation}
\dot{b}(k)=k\;u(k)b(k) - \eta k^2 b(k)
\end{equation}
leading to the following growthrate
\begin{equation}
\gamma(k)=k\;u(k) - \eta k^2.
\label{growthrate}
\end{equation}
Assuming a Kolmogorov spectrum in the flow inertial range, $u(k)= \varepsilon^{1/3}k^{-1/3}$, we find
$\gamma(k)=\varepsilon^{1/3}k^{2/3}-\eta k^2$ which is plotted in figure \ref{gamma}.
In the kinematic regime due to the linearity of the induction equation
all magnetic scales grow with the same growthrate $\gamma_{kin}=\max \gamma(k)$.
It occurs at $k_{kin} = 3^{-3/4}\varepsilon ^{1/4}\eta^{-3/4}$ with the value $\gamma_{kin} = 2/(3\sqrt{3})\varepsilon^{1/2}\eta^{-1/2}$.
The resistive scale $k_{\eta}^{-1}$ is the scale for which the shear $ku(k)$ is compensated by the resistive dissipation $\eta k^2$, or equivalently it is the scale for which $\gamma(k)=0$. We have $k_{\eta}=\varepsilon ^{1/4}\eta^{-3/4}$ and $k_{kin} =0.438\;k_{\eta}$.

Defining a magnetic Reynolds number at wave number $k$ by $Rm(k)=k^{-1}u(k)/\eta$, we have $Rm(k)=\varepsilon^{1/3}k^{-4/3}/\eta$ in the inertial range and consequently $Rm(k_{\eta})=1$ and $Rm(k_{kin})=3$.
Comparatively the magnetic Reynolds number at the forcing scale $k_F^{-1}$ is much larger,
scaling as $Rm(k_F)\sim Rm(k_{kin})/\eta$. Surprisingly the kinematic growthrate is given by the small scale $k_{kin}^{-1}$ for which $Rm(k)$ is of order unity
and not by the forcing scale $k_F^{-1}$ for which $Rm(k)$ is much larger.
In addition $k_{kin}$ and $\gamma_{kin}$ are viscosity independent in the limit of low $Pm$, suggesting that additional scales in the kinetic inertial range do not change the result. This is also consistent with the fact that the dynamo threshold should not depend on $Pm$ in the limit of low $Pm$ as shown previously \cite{Ponty05,Stepanov06, Schekochihin07}.

Coming back to the shell model, as the induction is linear in $B=(b_1, b_2, \cdots, b_N)$ where $N$ is the maximum number of shells, we end up with a system of the form $dB/dt = {\cal M} \cdot B$ where ${\cal M}$ is a $N$ by $N$ matrix depending on the flow $U=(u_1, u_2, \cdots, u_N)$. During the kinematic regime the flow $U$ is statistically stationary (the kinematic regime being defined as long as the Lorentz forces are negligible), implying that the system $dB/dt = {\cal M} \cdot B$ can be solved as an eigenvalue problem. Among the $N$ eigenvalues of the matrix ${\cal M}$, the largest one should correspond to $\gamma_{kin}$ (provided that our phenomenological approach is correct). In addition the corresponding eigenvector should be peaked at $k_{kin}$.
\begin{figure}
\begin{tabular}{@{\hspace{0cm}}c@{\hspace{0cm}}c@{\hspace{0cm}}l@{}}
    \includegraphics[width=0.5\textwidth]{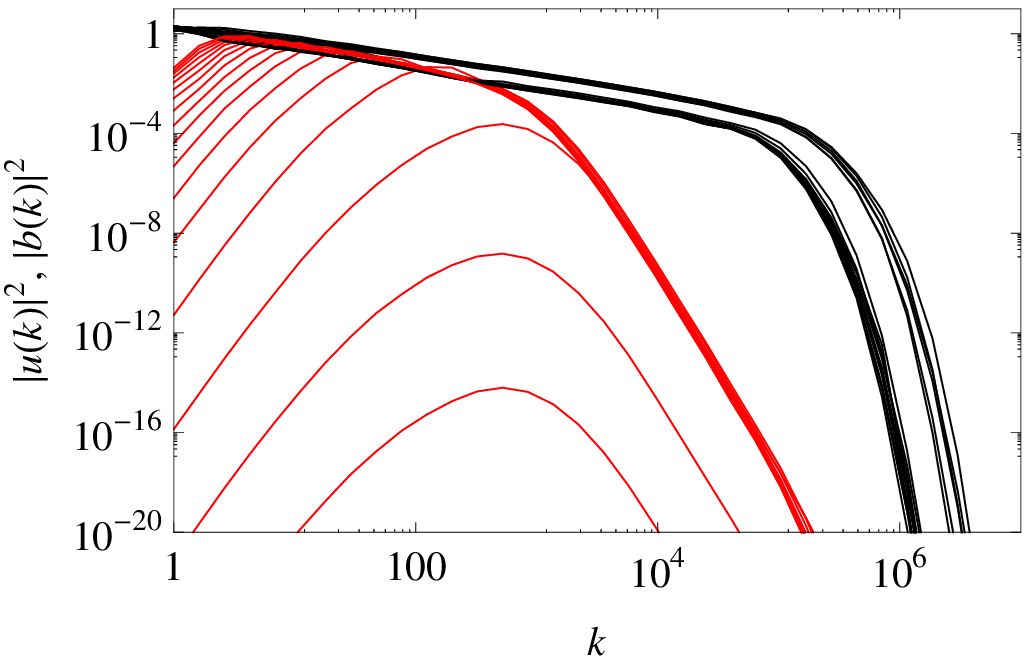}
    &
\includegraphics[width=0.5\textwidth]{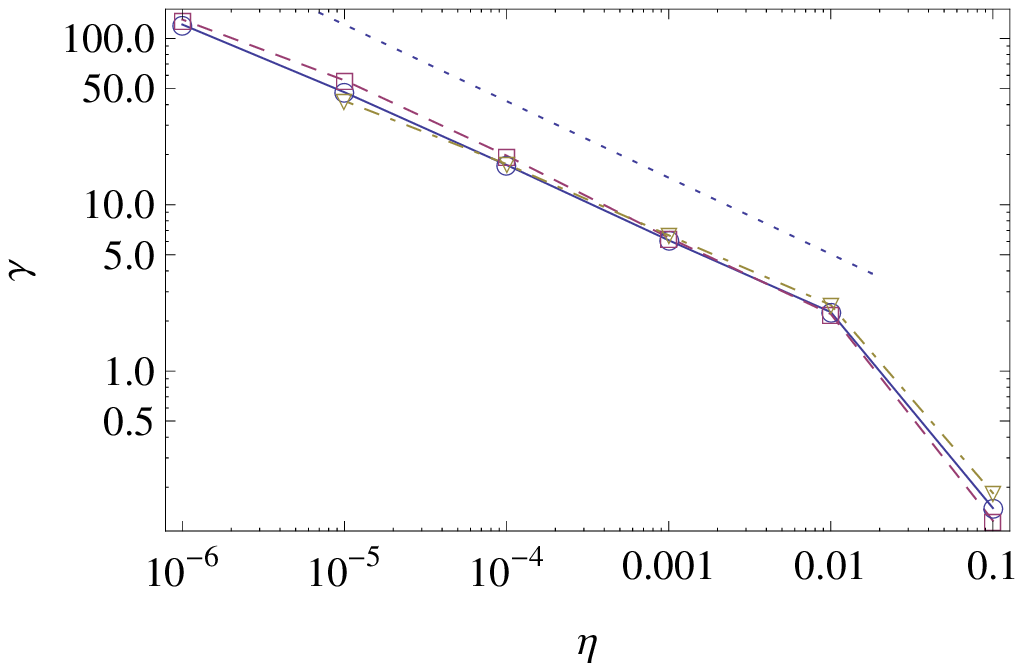}
    \\*[-0.5cm]
   (a) & (b)\\*[0.cm]
\includegraphics[width=0.5\textwidth]{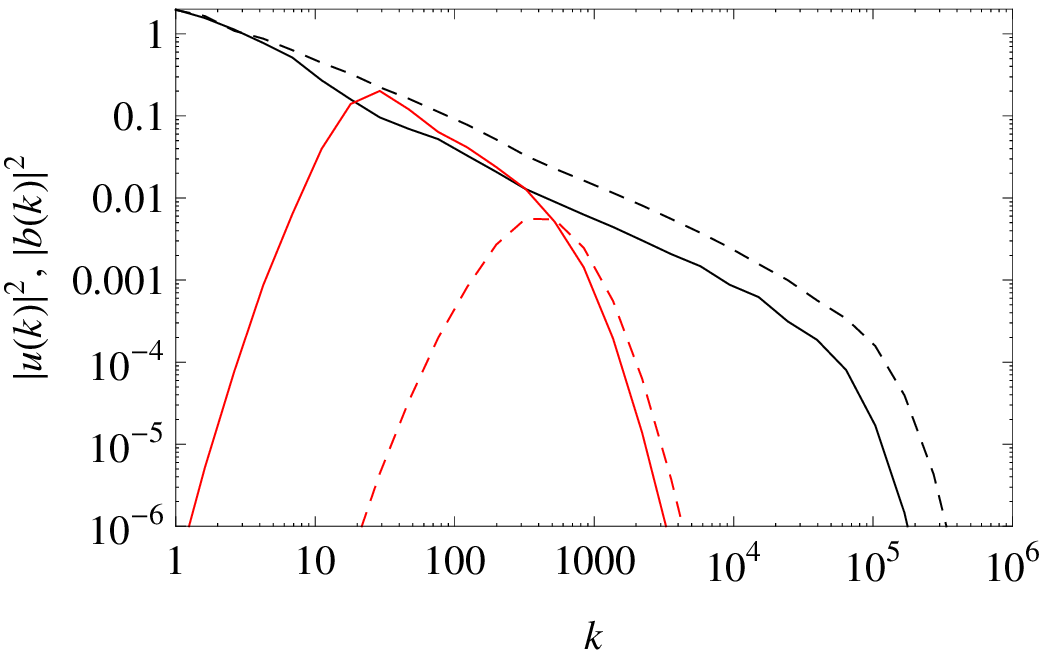}
&
    \includegraphics[width=0.5\textwidth]{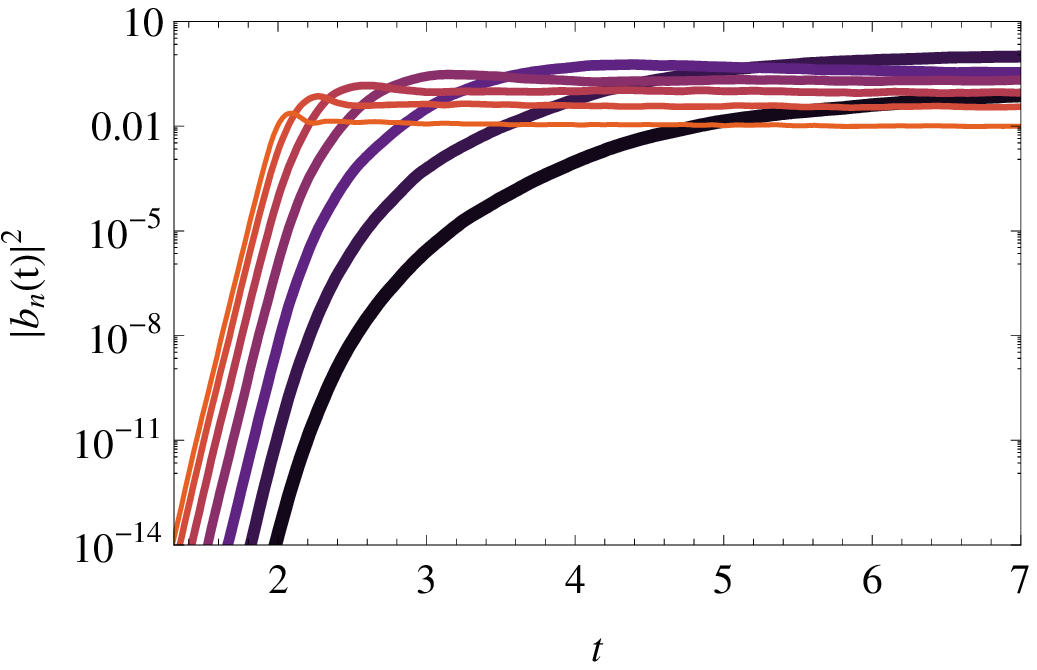}
    \\*[-0.5cm]
   (c) & (d)\\*[0cm]
\includegraphics[width=0.5\textwidth]{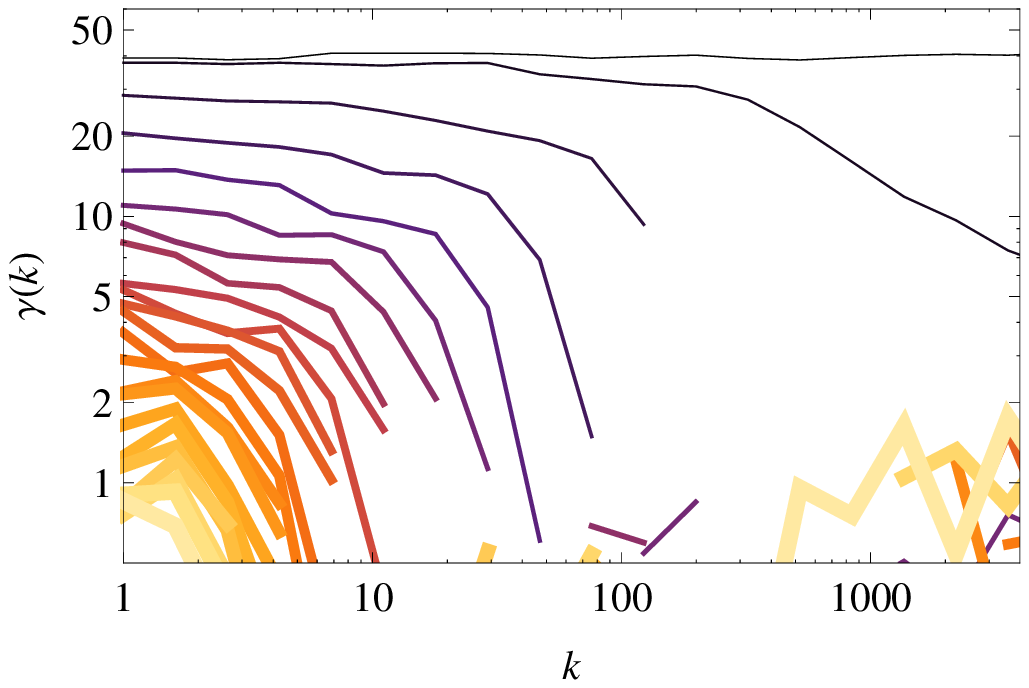}
&
    \includegraphics[width=0.5\textwidth]{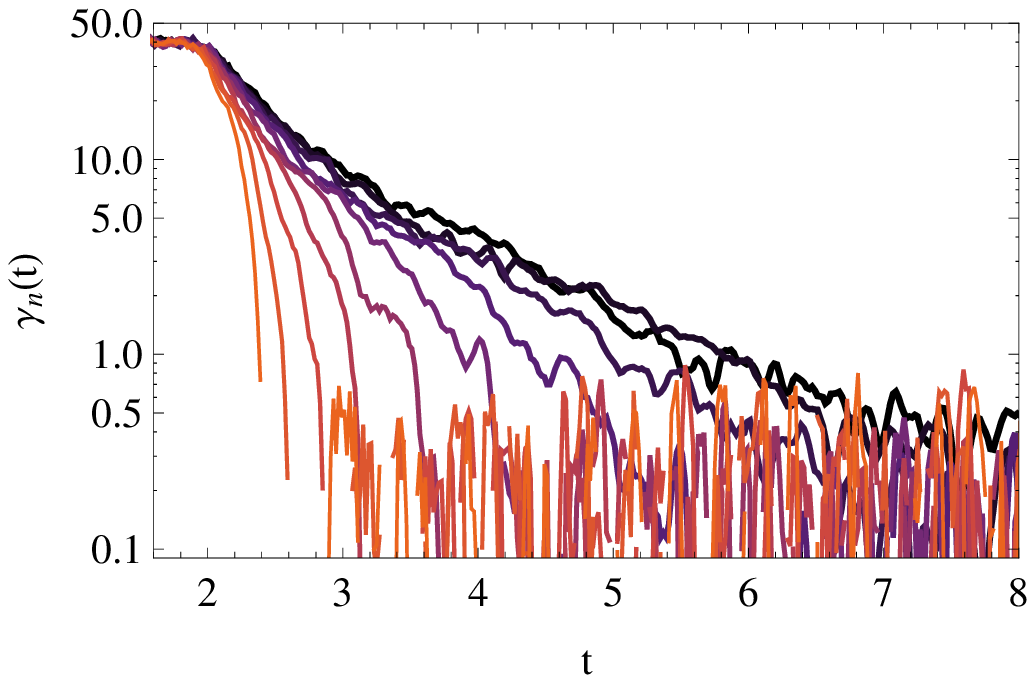}
    \\*[-0.5cm]
   (e) & (f)\\*[0cm]
  \end{tabular}
\caption{Low magnetic Prandtl number regime $Pm \ll 1$. In all figures except (b), $\nu=10^{-7}$ and $\eta=10^{-4}$. In (a) and (c), $|u(k,t)|^2$ (black curves) and $|b(k,t)|^2$ (red curves) are plotted versus $k$ at different times. In (d) $|b(k,t)|^2$ is plotted versus time for several values of $k$. The thickest (and darkest) curve corresponds to the largest scale. The magnetic energy growthrate $\gamma(k,t)$ is plotted versus $k$ in (e) and versus $t$ in (f). In (e) the curves evolve from top to bottom along time. In (f) the color scale is identical to (d). In (b) the growthrate during the kinematic regime is plotted versus $\eta$ for several viscosities.}
\label{figlowpm}
\end{figure}

In figure \ref{figlowpm}a $|u(k)|^2$ (black curves) and $|b(k)|^2$ (red curves) are plotted versus k at different times, for $\nu=10^{-7}$ and $\eta=10^{-4}$.
The kinetic energy satisfies the Kolmogorov scaling $u^2=\varepsilon^{2/3}k^{-2/3}$ (some deviation occurs due to intermittency as shown in PS07).  In the kinematic regime the magnetic spectrum (red curves in \ref{figlowpm}a) is peaked at a scale in agreement with $k_{kin}$ and grows exponentially with the same growthrate at all scales as shown in figure \ref{figlowpm}d, with a growthrate in agreement with $\gamma_{kin}$.

The kinematic growthrate has been calculated for other values of $Pm < 1$. It is plotted versus $\eta$ in figure \ref{figlowpm}b for three viscosities ($\nu=10^{-7}, 10^{-5},10^{-3}$). It is clear from the overlapping curves that the growthrate does not depend on the viscosity. In addition the curves show a clear power dependence in $\eta$.
The above phenomenology tells us that $\gamma_{kin} \sim \eta^r$ with $r=-0.5$. Instead we find from figure \ref{figlowpm}b that $r=-0.46$. This apparent discrepancy can be explained from the fact that, as $u_n$ is highly intermittent, we have $u_n\propto k_n^{-\zeta}$ with $\zeta=0.369$ (instead of the Kolmogorov scaling $\zeta=1/3$). This new scaling implies $k_{kin}\sim \eta^{-1/(1+\zeta)}$ and $\gamma_{kin} \sim \eta^{-(1-\zeta)/(1+\zeta)}$, leading to $r \approx -0.46$.

\subsection{Dynamic regime}
\label{lowpmdynamic}
 The kinematic regime ends when the saturation starts. Here this corresponds to the time when the magnetic energy at wave number $k_{kin}$ becomes comparable to the kinetic one $b(k_{kin}) \approx u(k_{kin})$.  Then due to the Lorentz forces the magnetic energy at that scale saturates and its growthrate decreases to zero. Meanwhile, the magnetic energy of the next smaller wave number ($k_{kin}/\lambda$ in the shell model) becomes the one with the largest growthrate. Then the wave numbers $k\le k_{kin}/\lambda$ continue to grow with the new growthrate $\gamma(k_{kin}/\lambda)$ until $b(k_{kin}/\lambda)$ saturates $b(k_{kin}/\lambda)\approx u(k_{kin}/\lambda)$.  Then the next smaller wave number $k_{kin}/\lambda^2$ becomes the one with the largest growthrate and successively all the next smaller wave numbers change their growthrate in accordance with the growthrate spectrum $\gamma(k)$ given in figure \ref{gamma}. This saturation mechanism lasts until the smallest wave number is saturated. Then equipartition between both kinetic and magnetic energies is obtained at all scales.

Such a scenario is quantitatively illustrated by the mode crossing in figure \ref{figlowpm}(d) where the magnetic energy is plotted versus time for several wave numbers. The way the magnetic wave numbers get their energy is reversed when compared to the direct hydrodynamic cascade, because  it starts at $k_{kin}$ towards smaller wave numbers. We speak of an inverse `` cascade '' \cite{Pouquet76}.
It is also illustrated in figure \ref{figlowpm}(e) where the magnetic growthrate $\dot{b}(k,t)/b(k,t)$ is plotted versus $k$ at different times. In the kinematic regime the growthrate is the same at all wave numbers (top black curve). Then as time goes, the plateau of constant growthrate decreases in both intensity and range of involved wave numbers.

An interesting feature of the route to saturation is shown in figure \ref{figlowpm}(c).
The injection rate of kinetic energy $\varepsilon$ partly dissipates at the resistive scale $k_{\eta}^{-1}$.
A direct consequence is that the viscous scale $k_{\nu}^{-1}$ shifts to larger values during the saturation regime.

\begin{figure}
\begin{tabular}{@{\hspace{0cm}}c@{}}
\includegraphics[width=1\textwidth]{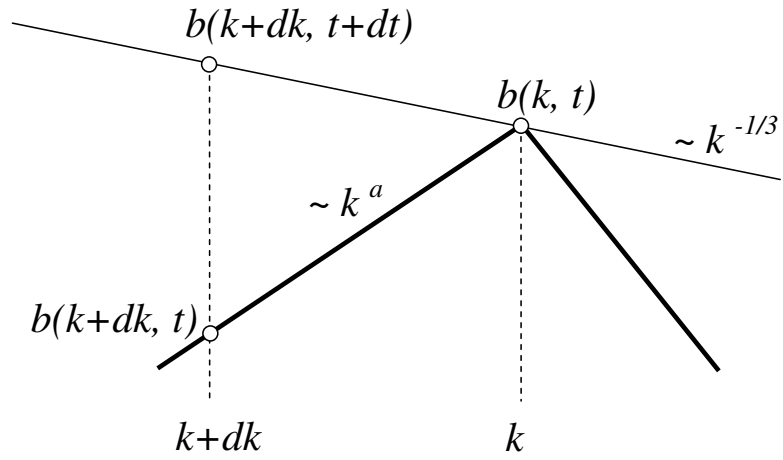} \\*[-14cm]
\end{tabular}
\caption{Scale by scale saturation mechanism.  }
\label{saturation}
\end{figure}

Finally in figure \ref{figlowpm}(f) the growthrate $\dot{b}(k,t)/b(k,t)$ is plotted versus time for each scale $k$. A signature of the saturation scenario described above is the time evolution of the growthrate at the largest scale (black curve). We shall compare it to the following phenomenological description which is sketched
schematically in figure \ref{saturation}.

\begin{figure}
\includegraphics[width=0.5\textwidth]{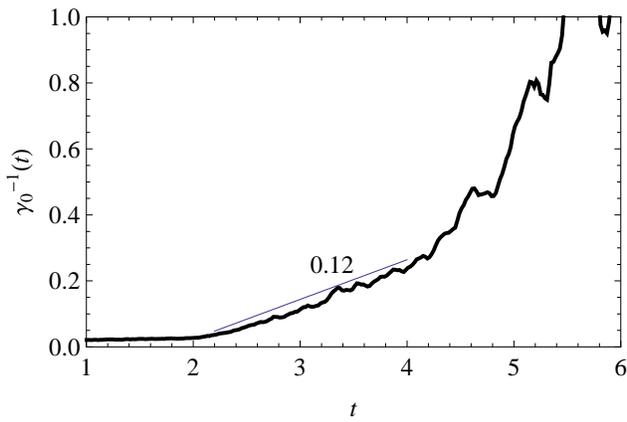}
\caption{The largest scale inverse growthrate $\gamma^{-1}(k=1)$ versus $t$ for $Pm=10^{-3}$.}
\label{slopelowpm}
\end{figure}

At time $t$ the magnetic field at wave number $k$ reaches its saturation value $b(k,t)=u(k)$.
The left part of the magnetic spectrum scaling as $k^a$, we have $b(k+dk,t)=b(k,t) (1+dk/k)^a$.
Then for $t\le t' \le t+dt$ the magnetic field at scale $k+dk$
grows exponentially as
$b(k+dk,t') = b(k+dk,t) \exp(\gamma(k+dk)\cdot (t'-t))$. At time $t'=t+dt$
it reaches its saturation value
$b(k+dk,t+dt)=u(k+dk)$.
For a Kolmogorov turbulence $u(k+dk)=u(k) (1+dk/k)^{-1/3}$ and assuming that $dk^2 + dt^2 \ll 1$, we finally find
\begin{equation}
dk/dt=-\frac{k\gamma(k)}{(a+1/3)}.
\end{equation}
From (\ref{growthrate}) and for $k\le k_{kin}$, a rough approximation of $\gamma(k)$ is $\gamma(k)\sim \gamma_{kin}(k/k_{kin})^{2/3}$ leading to
\begin{equation}
\frac{d k}{d t} = - \frac{\gamma_{kin}}{(a+1/3)k_{kin}^{2/3}}k^{5/3}.
\end{equation}
Defining $t_0$ as the time at which the saturation starts (at $k=k_{kin}$), we find
that the magnetic scale which saturates at subsequent time $t>t_0$ satisfies
\begin{equation}
\frac{k(t)}{k_{kin}} = \left(1 + \frac{2 \gamma_{kin}}{(1+3a)}(t-t_0)\right)^{-3/2}.
\end{equation}
The growthrate of the largest magnetic scale then satisfies
\begin{equation}
\gamma^{-1}(t)=\gamma_{kin}^{-1} + \frac{2}{(1+3a)}(t-t_0).
\label{gamma2}
\end{equation}
This is the signature of the saturation scenario that we want to compare to the shell model results.
In the shell model we have $a \approx 2$ (corresponding to an `` infrared'' spectrum $b^2(k)\sim k^4$ as found
e.g. in Christensson etal. 2001).
Therefore we expect the inverse of the magnetic energy growthrate in shell $n=0$ to satisfy
$(2\gamma)^{-1} = (2\gamma_{kin})^{-1} + (t-t_0)/7$. In addition intermittency again changes slightly the result. Indeed for $u \sim k^{-\zeta}$, we find that
\begin{equation}
\gamma^{-1}(t)=\gamma_{kin}^{-1} + \frac{1-\zeta}{(a+\zeta)}(t-t_0)
\label{gamma3}
\end{equation}
leading to a slope $d(2\gamma)^{-1}/dt=0.133$ instead of 1/7 (without intermittency).
In figure \ref{slopelowpm}, $(2\gamma)^{-1}$ is plotted versus time. We find indeed a good agreement with the previous expectation, at least for $9<t<11$. At subsequent times the curve diverges because all magnetic wave numbers are saturated expect the 2 smallest ones ($n=0,1)$. As the turbulence forcing is applied at these wave numbers, they do not belong to the inertial range and then the previous picture fails. Therefore the saturation scenario works until the saturation scale becomes close to the forcing scale.

\section{High $Pm$ dynamo action}
\label{highpm}
\subsection{Kinematic regime}

At high $Pm$, the resistive scale $k_{\eta}^{-1}$ is much smaller than the viscous scale $k_{\nu}^{-1}$. The flow scale which is responsible for the magnetic growth in the kinematic regime is the one which has the highest shear $ku(k)$. Assuming a Kolmogorov spectrum in the inertial range, we have $ku(k)=\varepsilon^{1/3}k^{2/3}$ which is maximum at the viscous scale $k_{\nu}^{-1}$ (at smaller scales the viscous dissipative range starts and $u(k)$ falls down).
Then the magnetic growthrate in the kinematic regime can be estimated as \cite{Schekochihin02a,Schekochihin04b}
\begin{equation}
\gamma = k_{\nu} u_{\nu} = (\varepsilon/\nu)^{1/2}.
\label{gammahighpm}
\end{equation}
We note that $\gamma$ depends on the viscosity but not on the diffusivity.
The corresponding phenomenologic induction equation is
\begin{equation}
\dot{b}(k)=k_{\nu}u_{\nu}b(k) - \eta k^2 b(k) \quad \quad \mbox{for} \quad \quad k>k_{\nu}
\label{induchighpm}
\end{equation}

The resistive scale $k_{\eta}^{-1}$ is the scale for which the shear $k_{\nu}u_{\nu}$ compensates for the resistive dissipation $\eta k^2$. As at the viscous scale the shear also compensates for the viscous dissipation, we have $\eta k_{\eta}^2=\nu k_{\nu}^2$, leading to $k_{\eta}/k_{\nu}=Pm^{1/2}$.

From (\ref{induchighpm}) we expect a flat spectrum of $|b(k)|^2$ for $k_{\nu} < k < k_{\eta}$.
In addition the ``infrared'' spectrum $k < k_{\nu}$ is enslaved to the growth of the largest wave numbers $k>k_{\nu}$.
\begin{figure}
\begin{tabular}{@{\hspace{0cm}}c@{\hspace{0cm}}c@{\hspace{0cm}}l@{}}
    \includegraphics[width=0.5\textwidth]{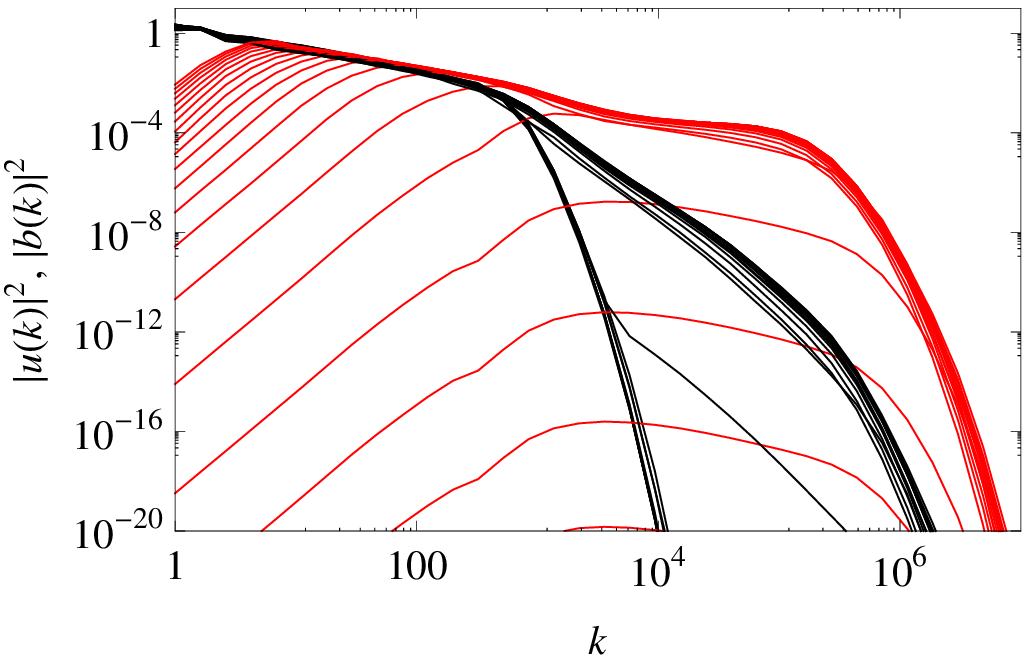}
    &
\includegraphics[width=0.5\textwidth]{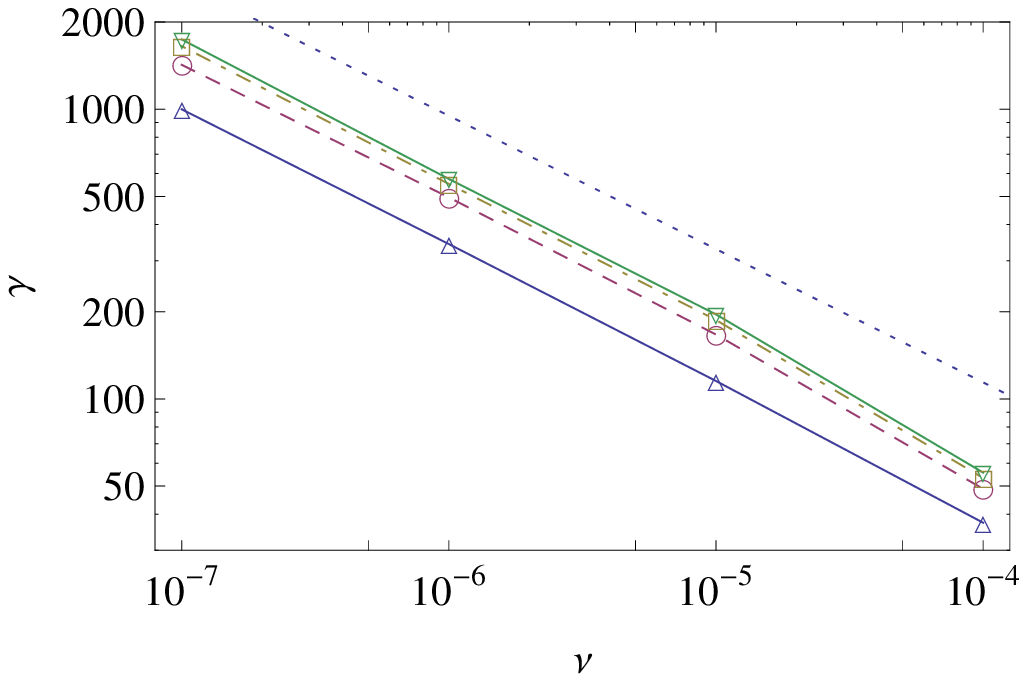}
    \\*[-0.5cm]
   (a) & (b)\\*[0.5cm]
\includegraphics[width=0.5\textwidth]{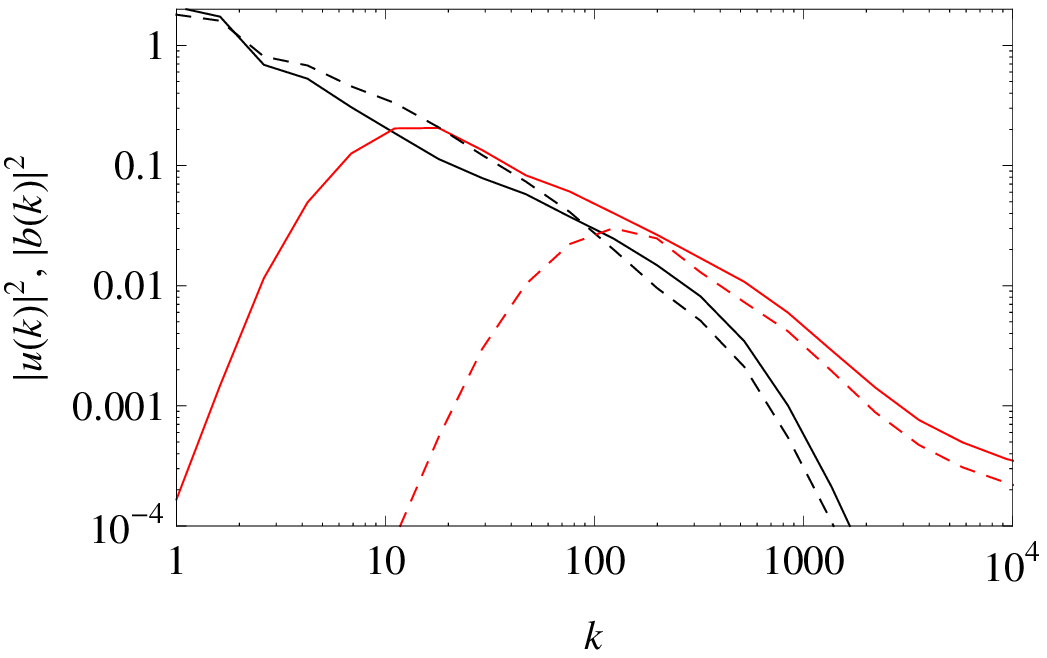}
&
    \includegraphics[width=0.5\textwidth]{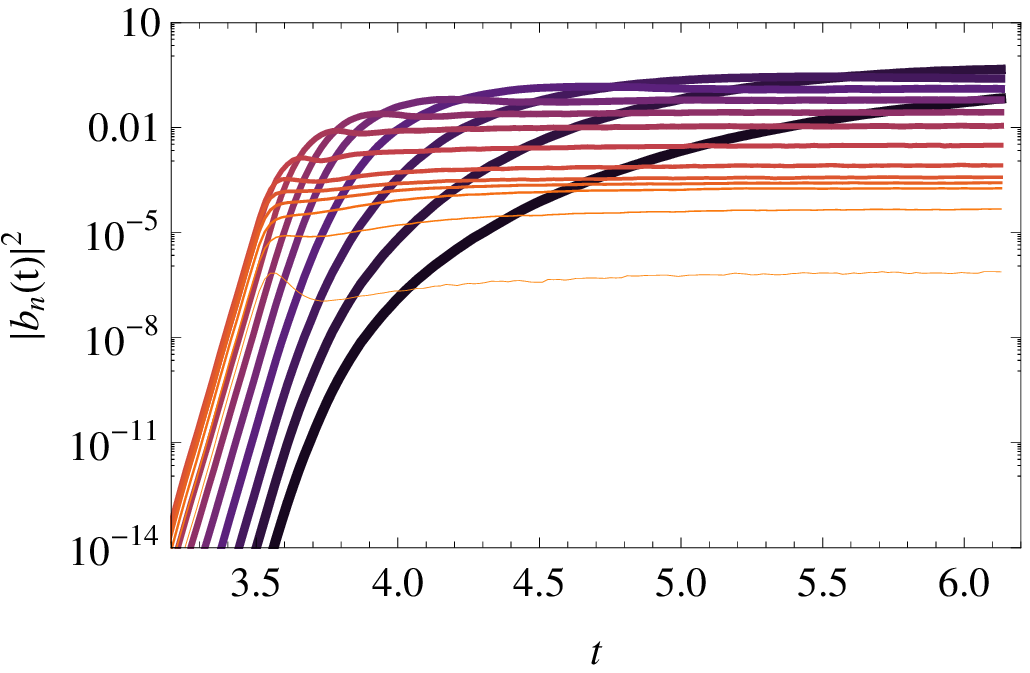}
    \\*[-0.5cm]
   (c) & (d)\\*[0cm]
\includegraphics[width=0.5\textwidth]{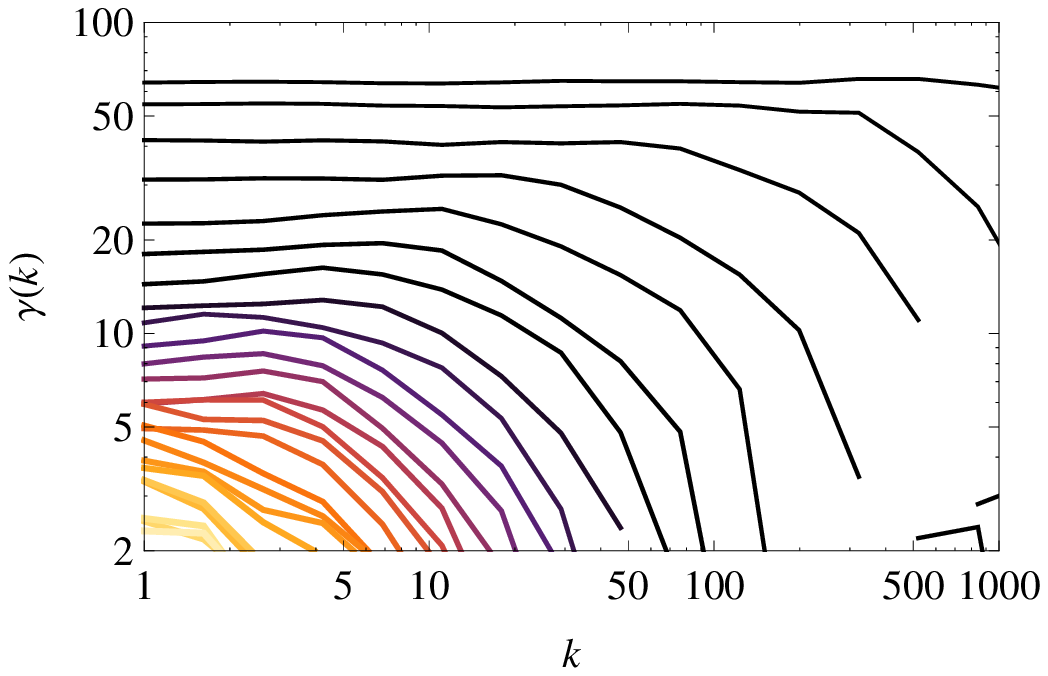}
&
    \includegraphics[width=0.5\textwidth]{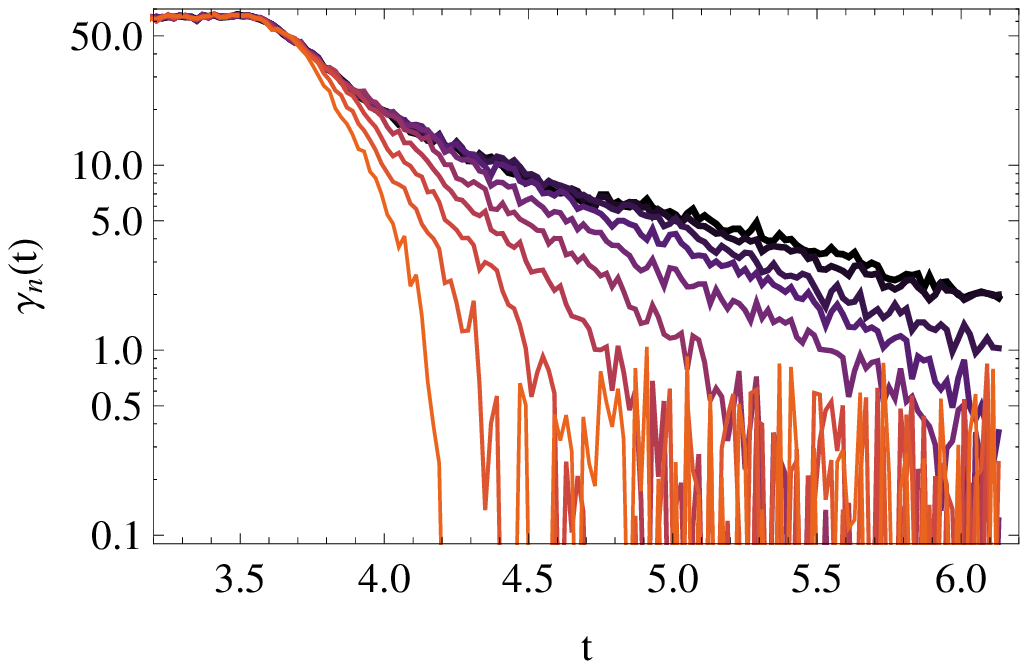}
    \\*[-0.5cm]
       (e) & (f)\\*[0cm]
  \end{tabular}
\caption{High magnetic Prandtl number regime $Pm \ll 1$. In all figures except (b), $\nu=10^{-4}$ and $\eta=10^{-8}$. The caption is the same as in figure \ref{figlowpm} except (b) for which the growthrate during the kinematic regime is plotted versus $\nu$ for several diffusivities.
}
\label{fighighpm}
\end{figure}

Comparing (\ref{eq_b}) with (\ref{induchighpm}) we can show that the only way to account for the non local term
$k_{\nu}u_{\nu}b(k)$ is to take $\alpha=-1$ in the shell model.
In figure \ref{fighighpm}(a), $|u(k)|^2$ (black) and $|b(k)|^2$ (red) are plotted versus $k$ at different times, for $\nu=10^{-4}$ and $\eta=10^{-8}$.
In the kinematic regime, the magnetic spectrum is not completely flat for $k_{\nu}< k < k_{\eta}$. This is probably an effect of the additional non local energy transfers involving kinetic scales larger than $k_{\nu}^{-1}$.

The kinematic growthrate that we find agrees with (\ref{gammahighpm}).
It has been calculated for other values of $Pm > 1$. It is plotted versus $\nu$ in figure \ref{fighighpm}(b) for four values of the magnetic Prandtl number ($Pm=10, 10^2,10^3,10^4$). For increasing values of $Pm$ the curves shift from bottom to top and reach an asymptotic limit showing that the growthrate becomes $\eta$-independent at high $Pm$. In addition the curves show a clear power scaling in $\nu$.
The above phenomenology tells us that $\gamma_{kin} \sim \nu^r$ with $r=-0.5$. Instead we find from figure \ref{fighighpm}(b) that $r=-0.46$. As in the low $Pm$ case, this apparent discrepancy results from the fact that the velocity $u_n$ is intermittent.
Then for $u_n\propto k_n^{-\zeta}$ with $\zeta=0.369$, we find $k_{\nu}\sim \nu^{-1/(1+\zeta)}$ and $\gamma_{kin} \sim \nu^{-(1-\zeta)/(1+\zeta)}$, leading to $r \approx -0.46$.

\subsection{Dynamic regime}
As explained before, during the kinematic regime, the magnetic field $b(k)$ at wave number $k>k_{\nu}$ receives energy from $u_{\nu}$. It also releases energy to $u(k)$ locally such that the Lorentz force partially compensates for the viscous dissipation. This is visible in figure \ref{fighighpm}(a) where in the viscous range for scales $k_{\nu} < k < k_{\eta}$, the slope of $|u(k)|^2$ becomes less steep as soon as the magnetic energy is sufficiently large.

At wave number $k=k_{\nu}$ when $|b(k_{\nu})|^2 \approx |u(k_{\nu})|^2$ the saturation starts and again the saturation occurs from the smaller to larger scales as shown in figure \ref{fighighpm} (c) to (e). Again the inverse ``cascade'' saturation scenario seems to be the relevant one.

\begin{figure}
\includegraphics[width=0.5\textwidth]{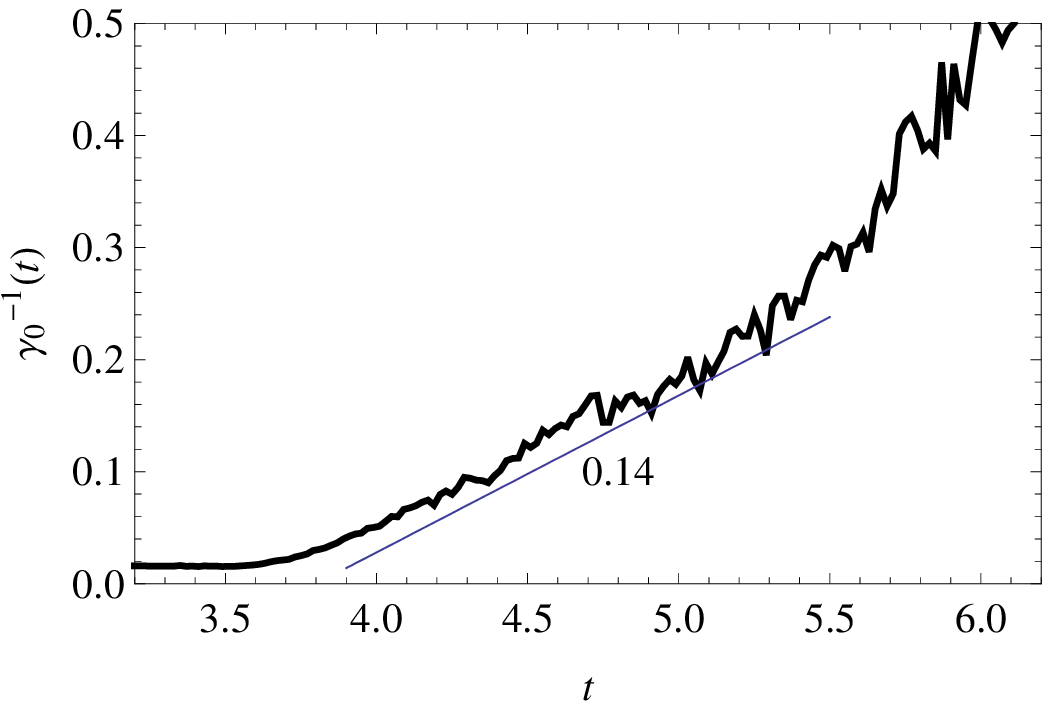}
\caption{The largest scale inverse growthrate $\gamma^{-1}(k=1)$ versus $t$ for $Pm=10^4$.}
\label{slopehighpm}
\end{figure}

In figure \ref{slopehighpm}, $(2\gamma)^{-1}$ is plotted versus time. For $3.5<t<7.7$ we find a good agreement with the phenomenological prediction derived in section \ref{lowpmdynamic}. As explained previously, at  subsequent times the curve diverges as the remaining scales to saturate are larger than the inertial range.

\section{Discussion}
In the kinematic regime, we found that the magnetic growthrate satisfies
$\gamma \sim (\varepsilon / \eta)^{1/2}$ for $Pm\ll 1$ and $\gamma \sim (\varepsilon / \nu)^{1/2}$ for $Pm\gg 1$ and is therefore always fast \cite{Childress95}.
In both cases this corresponds to a small-scale dynamo (or fluctuation dynamo in the terminology of Schekochihin \textit{etal.} 2007) in opposition to a large-scale (or mean-field dynamo).
At low $Pm$ the magnetic spectrum is peaked at $k=k_{kin} \approx 0.4 k_{\eta}$ and the energy transfer
from kinetic to magnetic is mainly local (as shown in PS07 the non local transfers from small kinetic to large magnetic scales is less than 20 $\%$ than the local transfers). At high $Pm$ the magnetic spectrum is almost flat between the viscous and resistive scale and the energy transfer is mainly non local from $u(k_{\nu})$
to $b(k)$ with $k_{\nu}<k<k_{\eta}$.
In the dynamic regime we found an inverse ``cascade'' mechanism which explains the route to saturation. The relevance of this mechanism does not depend on $Pm$. When all magnetic scales are saturated the resulting equilibrium between kinetic and magnetic energy is not equipartition with a slight excess of magnetic energy \cite{Haugen03}.
We find that the ratio $|b(k)/u(k)|$ is constant for $k_F<k<\min\{{k_{\nu},k_{\eta}}\}$ corresponding to a residual energy with a Kolmogorov scaling \cite{Muller05}.

Dynamo action is usually found in natural objects with strong rotation. A transition from a Kolmogorov spectrum $E(k)\sim k^{-5/3}$ for weak rotation to $E(k)\sim k^{-2}$ for strong rotation is predicted by weak turbulence theory, and observed in experiments \cite{Baroud02,Baroud03} and numerical simulations \cite{Smith99} including shell models \cite{Hattori04}. It is then of interest to see how the previous phenomenology is changed.
At low $Pm$ for $u(k)\sim k^{-\zeta}$ with $1/3\le \zeta \le 1/2$, we derive the quantities given in table \ref{summary} and give the numerical values expected for the three scalings, Kolmogorov, intermittent and strongly rotating.
The change of the power scalings from weak to strong rotation should be sufficient to be identified in measurements or numerical calculations. On the other hand $k_{kin}/k_{\eta}$ and $Rm(k_{kin})$ do not change much.
\begin{table}
\begin{tabular}{@{\hspace{0.8cm}}r@{\hspace{0.8cm}}c@{\hspace{0.8cm}}c@{\hspace{0.8cm}}c@{\hspace{0.8cm}}c@{}}
    &Phenomenology&Kolmogorov&Intermittency&Strong rotation\\*[0cm]
    $\zeta=$&             &$1/3$&$0.369$&$0.5$\\
    $k_{kin}/k_{\eta}=$ & $\left(\frac{1-\zeta}{2}\right)^{\frac{1}{1+\zeta}}$&0.439&0.431&0.397\\
    $Rm(k_{kin})=$ & $\frac{2}{1-\zeta}$ & 3 & 3.17 & 4 \\
    $k_{kin},k_{\eta} \sim$ & $\eta^{-\frac{1}{1+\zeta}}$ & $\eta^{-0.75}$ & $\eta^{-0.73}$ & $\eta^{-0.66}$ \\
        $k_{\nu} \sim$ & $ \nu^{-\frac{1}{1+\zeta}}$ & $\nu^{-0.75}$ & $\nu^{-0.73}$ & $\nu^{-0.66}$\\
            $\gamma_{kin} \sim$ & $ \eta^{-\frac{1-\zeta}{1+\zeta}}$ & $\eta^{-0.5}$ & $\eta^{-0.46}$ & $\eta^{-0.33}$ \\
            $\frac{d(2\gamma)^{-1}}{dt} \sim$ & $  \frac{1-\zeta}{a+\zeta}$&0.143&0.133&0.1
    \end{tabular}
    \caption{Summary of the numerical values expected from phenomenological predictions at low $Pm$ for different velocity spectra $u(k)\sim k^{-\zeta}$.}
    \label{summary}
\end{table}
\acknowledgements{
Most of this work was done during the Summer Program on MHD Turbulence at the Universite Libre de Bruxelles during July 2007. The organizers D. Carati and B. Knaepen are warmly thanked as well as the sponsors of this program.
R.Stepanov thanks for financial support from RFBR grants (07-01-96007 ural and 07-02-00127) and the Russian
Federation President grant MK-4338.2007.1. The simulations were performed on the computer cluster of IMM (Ekaterinburg, Russia).}
\newpage
\bibliographystyle{apj}

\end{document}